\documentclass{ws-ijmpa}
\usepackage[super,compress]{cite}
\begin{document}

\markboth{Abdulla Abdulsalam and Prashant Shukla}
{Suppression of bottomonia states in finite size quark gluon plasma 
in PbPb collisions at Large Hadron Collider}

\catchline{}{}{}{}{}

\title{Suppression of bottomonia states in finite size quark gluon plasma 
in PbPb collisions at Large Hadron Collider}

\author{Abdulla Abdulsalam and Prashant Shukla$^*$}
\address{Nuclear Physics Division, Bhabha Atomic Research Center, Mumbai, 400085, India \\
$^*$pshukla@barc.gov.in}

\maketitle

\begin{history}
\end{history}

\begin{abstract} 
  The bottomonium states due to their varying binding energies dissolve at different temperatures and 
thus their nuclear modification factors and relative yields have potential to map the properties of 
Quark Gluon Plasma (QGP). We estimate the suppression of bottomonia states due to color screening 
in an expanding QGP of finite lifetime and size with the conditions relevant for PbPb collisions at LHC. 
The properties of $\Upsilon$ states and recent results on their dissociation temperatures 
have been used as ingredients in the study. The nuclear modification factors and the ratios of 
yields of $\Upsilon$ states are then obtained as a function of transverse momentum and centrality. 
We compare our theoretical calculations with the bottomonia yields measured with CMS in 
PbPb collisions at $\sqrt{s_{\rm NN}}$ = 2.76 TeV. The model calculations explain the data very well.

\end{abstract} 

\ccode{PACS numbers: 12.38.Mh, 24.85.+p, 25.75.-q}

\section{Introduction}
\label{Intro}
 The heavy ion collisions produce matter at extreme temperatures and densities where 
it is expected to be in the form of Quark Gluon Plasma 
(QGP), a phase in which the quarks and gluons can move far beyond  the size of a nucleon 
making color degrees of freedom dominant in the medium. 
  The experimental effort to produce such matter started with low energy CERN accelerator 
SPS and evolved through voluminous results 
from heavy ion collision at Relativistic Heavy Ion Collider (RHIC) \cite{INTRO_Arsene, INTRO_Back, INTRO_Adams, INTRO_Adcox}.
The recent results from Large Hadron Collider (LHC) experiments \cite{QGP_Tc} are 
pointing towards formation of high temperature system in many ways similar to the matter
produced at RHIC. 
  One of the most important signal of QGP is the suppression of 
quarkonium states \cite{SATZ}, both of the charmonium ($J/\psi$, $\psi(2S)$, $\chi_{c}$, etc) 
and the bottomonium ($\Upsilon(1S)$ , $\Upsilon(2S)$, $\chi_{b}$, etc) families. This is thought to be a 
direct effect of deconfinement, when the binding potential between the constituents of a quarkonium state, 
a heavy quark and its antiquark, is screened by the colour charges of the surrounding light quarks and gluons. 
 The ATLAS and CMS experiments have carried out detailed quarkonia measurements in PbPb collisions 
with the higher energy and luminosity available at the LHC.
 The ATLAS measurements \cite{ATLAS} show suppression of inclusive $J/\psi$ with high transverse momenta $p_T$  
in central PbPb collisions compared to peripheral collisions at $\sqrt s_{NN} = 2.76$ TeV. 
  Similarly, CMS measured a steady and smooth decrease of suppression 
of prompt $J/\psi$ as a function of centrality with nuclear modification factor $R_{\rm AA}$ remaining $<$ 1 even 
in the peripheral bin \cite{JCMS}. 

 The melting temperature of the quarkonia states depends on their binding energy. The ground states, 
$J/\psi$ and $\Upsilon(1S)$ are expected to dissolve at significantly higher temperatures than the 
more loosely bound excited states. The difference in binding energies among different quarkonia indicate that
they melt in a hot QGP at different temperatures and the quarkonium spectrum can
serve as plasma thermometer \cite{SATZ2,Mocsy_Strik}.
  The $\Upsilon(2S)$ and $\Upsilon(3S)$ have smaller binding energies as compared to ground
state $\Upsilon(1S)$ and hence are expected to dissolve at a lower temperature. 
 With the 2011 PbPb run the CMS published results on sequential suppression of 
$\Upsilon(nS)$ states as a function of centrality \cite{CMSU2} with enlarged statistics
over their first measurement \cite{UCMS}
where a suppression of the excited $\Upsilon$ states with respect to the ground state have been observed  
in PbPb collisions compared to pp collisions at $\sqrt s_{NN} = 2.76$ TeV.

  The quarkonia yields in heavy ion collisions are also modified due to non-QGP effects such as
shadowing, an effect due to change of the parton distribution functions inside the nucleus,
and dissociation due to nuclear or comover interactions \cite{Vogt}. Due to higher mass, the 
nuclear suppression is expected to be less for bottomonia over charmonia.
  If large number of heavy quarks are produced in initial heavy ion collisions at LHC energy 
this could even lead to enhancement of quarkonia via statistical recombination \cite{Rapp1,Rapp2}. 
 The effect of regeneration is expected to be less significant for bottomonia as compared 
to charmonia since bottom quarks are much smaller in number as compared to charm quarks. 
 In addition, due to higher bottom mass the bound state properties obtained from 
potential models are more reliable. Thus recent years witness a shift in the 
interest to bottomonia. 
 The ratios of the yields of excited states to the ground states is considered 
even more robust QGP probe as the cold nuclear matter effects if any cancel out and can be 
neglected in the ratios. The calculation of ratios of $\Upsilon$ states was also made in 
few works e.g. \cite{UPsi_Blaiz,UPsi_Guni} in past which showed that the $p_T$ dependence 
of such ratio would show large variations and this would be a direct probe of the QGP. 
 
 In this paper, we calculate the bottomonia suppression due to color screening in an expanding
QGP using the model by Chu and Matsui \cite{CHU},
which takes into account the finite QGP lifetime and spatial extent. 
 We start by describing the properties of quarkonia obtained from potential models and then 
give a brief description of the model which is extended to get the survival 
probabilities of $\Upsilon$ states as a function of centrality of the collisions. 
 Finally we compare the model calculations with the experimental data recently 
measured by the CMS experiment.

\section{Properties of the $\Upsilon$ states from potential models}
Interaction between the heavy quark and its antiquark inside the quarkonium at zero temperature  
can be described by Cornell potential \cite{QPOT1, QPOT2,UPsi_KarschMehr}.
The solution of the Schrodinger equation for such potential gives mass, bound state radius and
the formation time $\tau_{F}$, the time needed to form a bound state after the production of heavy quark pairs.
 All parameters obtained with zero temperature 
potential using the parameter values given in \cite{UPsi_KarschMehr,QPROP} are summarized in 
first three rows of Table I, which describe well the experimentally 
observed quarkonia spectroscopy.          

The potential model can be extended to finite temperature with the main assumption that medium effects can be 
accounted for as a temperature-dependent potential. Instead of just looking at the individual
bound states (at $T$ = 0 where quarkonium is well defined), one could rather obtain a unified treatment of bound states, 
threshold and continuum by determining the spectral function. 
  Using a class of screened potentials based on lattice calculations of the static quark-antiquark free energy, 
spectral functions at finite temperature are calculated in a work \cite{UPsi_Mocsy2,UPsi_Mocsy3} and it was found that all 
quarkonium states, except the 1S bottomonium, dissolve in the deconfined phase at temperatures smaller than 1.5$T_C$.
An upper limit on binding energy and the thermal width of different quarkonia states are then estimated using  
spectral functions in the quark-gluon plasma. 
 Corresponding upper bounds on their dissociation temperatures $T_{D}$ \cite{UPsi_Mocsy3} are
given in second last row of Table I. We used slightly lower values of $T_{D}$ given in the last row 
to obtain a good match with measured $R_{\rm AA}$. 
\renewcommand{\arraystretch}{1.4}
\begin{table}[ph]
\tbl{Quarkonia properties from non-relativistic potential theory \cite{UPsi_KarschMehr,UPsi_Mocsy3}.}
{\begin{tabular}{@{}cccccc@{}} \toprule 
\hline\noalign{\smallskip}
  {\rm Bottomonium properties}        &    $\Upsilon(1S)$   & $\chi_b(1P)$  &  $\Upsilon(2S)$      & $\Upsilon(3S)$  &  $\chi_b(2P)$ \\
\hline\noalign{\smallskip}
 {\rm Mass~[GeV/$c^{2}$]}     &    9.46             &   9.99        &    10.02             & 10.36           &   10.26 \\
\hline
{Radius \rm [fm]}          &    0.28             &   0.44        &     0.56             & 0.78            &0.68    \\
\hline 
$\tau_{F}$ \rm [fm] \cite{UPsi_KarschMehr}  &    0.76             &   2.60        &     1.9              & 2.4             &     \\
\hline 
$T_D$ \rm [GeV] upper limit \cite{UPsi_Mocsy3}    &    2~$T_C$          &   1.3~$T_C$   &     1.2~$T_C$        & 1~$T_C$       &     \\
\hline
$T_D$ \rm [GeV] used in the present work  &    1.8~$T_C$          &   1.15~$T_C$   &     1.1~$T_C$        & 1.0~$T_C$       &     \\
\hline
\end{tabular} }
\label{prop}
\end{table}

\section{Quarkonia suppression in finite size QGP}
 The bottomonia survival probabilities due to color screening in an expanding QGP
are estimated using a dynamical model which takes into account 
the finite lifetime and spatial extent of the system \cite{CHU}. The competition between the resonance formation 
time $\tau_{F}$ and the plasma characteristics such as temperature, lifetime and spatial extent decide the 
$p_{T}$ dependence of the survival probabilities of $\Upsilon$ sates.  We describe the essential 
steps used to develop the model which is then extended to get 
the survival probabilities as a function of centrality of the collision.

  The model assumes that quark gluon plasma is formed at some initial entropy density 
$s_0$ corresponding to initial temperature $T_0$ at time $\tau_{0}$ which undergoes an 
isentropic expansion by Bjorken's hydrodynamics~\cite{UPsi_Bjork}. The plasma cools to an entropy density $s_D$ 
corresponding to the dissociation temperature $T_D$ in time $\tau_{D}$ which is given by 
\begin{equation}\label{bjork}
 \tau_{D} = \tau_0 \left( { s_0 \over s_D} \right) = \tau_0 \left( \frac{T_{0}}{T_{D}} \right)^{3},
\end{equation}
  As long as $\tau_{D}$/$\tau_{F}$ $>$ 1, quarkonium formation will be suppressed. 

 In the finite system produced in heavy ion collision, the suppression and entropy depend on the size 
of the system. The initial entropy density is assumed to be dependent on radius $R$ (decided by the
centrality of the collision) of the QGP \cite{CHU} as 
\begin{equation}\label{eprofile}
   s_0(r) = s_0 ~\left(1 - \left(\frac{r}{R}\right)^2\right)^{1/4},
\end{equation}
  Using Eq.~(\ref{bjork}) and Eq.~(\ref{eprofile}) one can obtain the $r$ dependence of $\tau_D$ as
\begin{eqnarray}\label{tDr}
   \tau_{D}(r) & = &  \tau_{D}(0)\left(1 - \left(\frac{r}{R}\right)^2\right)^{1/4}.
\end{eqnarray}
where $\tau_{D}(0)$ is the value of $\tau_{D}$ for resonances produced in the center of the system.

  Let a $Q\overline{Q}$ pair is created at the position ${\rm \bf r}$ in the transverse plane with a 
transverse momentum ${\rm \bf p_T}$ and transverse energy $E_{T}$ = $\sqrt{M^2 + p_{T}^2}$.
 The $\Upsilon$ formation time is $\tau_{F}\gamma$ which on equating with the screening duration $\tau_{D}(r)$  given 
in Eq~(\ref{tDr}) one obtains the critical radius $r_D$, which is the boundary of the suppression region as 
\begin{equation}
  r_D  = R\left(1 - \left(\frac{\gamma \tau_{F}}{\tau_{D}(0)}\right)^{4}\right)^{1/2}.
\end{equation}
where $\gamma$ = $E_T/M$ is the Lorentz factor associated with the transverse motion of the pair. 
 A bottom-quark pair can escape the screening region $r_D$ and form $\Upsilon$ if the position at 
which it is created satisfies
\begin{equation}\label{taumax}
| {\rm \bf r} + {\tau_{F} {\rm \bf p_{T}} \over M} | > r_D,
\end{equation}
where the screening region $r$ $<$ $r_D$ is shrinking because of the cooling of the system.
 Defining $\phi$ to be the angle between ${\rm \bf p_{T}}$  and ${\rm \bf r}$, the Eq.~(\ref{taumax}) leads to a range 
of $\phi$ for which the bottom-quark pair can escape:
\begin{equation}\label{cosmax}
   {\rm cos} \, \phi \ge z   ~~~~{\rm where } ~~~~ \nonumber \\ 
    z = \frac{ r_D^2 - r^2 - (\tau_{F}p_{T}/M)^2}{2r \,(\tau_{F}p_{T}/M)},
\end{equation}
  With this we can then calculate probability for the pair created at ${\rm \bf r}$ with transverse momentum ${\rm \bf p_T}$
to survive as
\begin{eqnarray}
\phi(r,p_{T})  &  =     1                 & \,\,\, z\le -1  \nonumber \\
               &  =  \left( { {\rm cos}^{-1}z \over \pi} \right)    & \,\,\, |z| < 1  \nonumber \\
               &  =    0                 & \,\,\, z\ge 1,   \nonumber  
\end{eqnarray}
  If the probability $\rho(r)$ of a quark pair to be created at $r$ which is symmetric in transverse plane 
is parameterized as
\begin{equation}
\rho(r) = \left(1 - \left( {r \over R} \right)^2\right)^{1/2},
\end{equation}
the survival probability of quarkonia becomes 
\begin{equation}
S(p_{T}, R) = \frac{\int_0^Rdr~r~\rho(r)~\phi(r,p_{T})}{\int_0^Rdr~r~\rho(r)}.
\end{equation}

The survival probability as a function of centrality can be obtained by integrating over 
$p_T$ as follows 
\begin{equation} 
  S(N_{\rm part})  = \int S(p_{T}, R(N_{\rm part}) ) \, Y(p_T) \,dp_T.
\end{equation}
Here Y($p_T$) is $p_T$ distribution (normalized to one) obtained from Pythia. 
The size $R = R(N_{\rm part})$ as a function of centrality is obtained in terms of the radius of the Pb 
nucleus given by $R_0 = r_0\, A^{1/3}$($r_0 = 1.2 \,$ fm) and the total number of participants $N_{\rm part0}=2A$ in head-on collisions as 
\begin{equation}\label{rnpart}
R(N_{\rm part}) = R_0 \, \sqrt{N_{\rm part} \over N_{\rm part0} }.
\end{equation}
We assumed initial temperature $T_0$ is the temperature in  0-5\% central collisions and calculated it
for a given initial time $\tau_0$ by
\begin{equation}\label{Int1}
T_{0}^{3}\tau_{0} = \frac{3.6}{4a_{q}\pi R_{0-5\%}^{2}}\left(\frac{dN}{d\eta}\right)_{0-5\%},
\end{equation}
Here $(dN/d\eta)_{0-5\%}$ = 1.5$\times$1600 obtained from the charge particle multiplicity measured in 
PbPb collisions at 2.76 TeV \cite{MULT} and $a_{q}$ = 37$\pi^{2}$/90 is the degrees of freedom we take in 
quark gluon phase. Using Eq.~(\ref{rnpart}) we can obtain the transverse size of the system
for 0-5\% centrality as $R_{0-5\%}$ = 0.92$R_0$. 
For $\tau_{0}$ = 0.1 fm/$c$, we obtain $T_{0}$ as 0.65 GeV using Eq.~(\ref{Int1}). 
The critical temperature is taken as $T_{C}$ = 0.170 GeV \cite{QGP_Tc}. 
The initial temperature as a function of centrality is calculated by 
\begin{equation}\label{Int2}
T(N_{\rm part})^3 = T_0^3 \, \left({dN/d\eta \over N_{\rm part}/2}\right) / \left({dN/d\eta \over N_{\rm part}/2}\right)_{0-5\%}.
\end{equation}
where $(dN/d\eta)$ is the multiplicity as a function of number of participants measured by ALICE experiment \cite{MULT}. 
Both ALICE and CMS \cite{CMSmult} measurements on multiplicity agree well with each other.
  Equation~(\ref{Int2}) giving the variation of initial temperature as a function of centrality 
differs from the approach taken in the work of Ref.~\cite{STR1} where it is taken to vary
as a third root of number of participants.
\begin{figure}
\begin{center}
\includegraphics[width=0.70\textwidth]{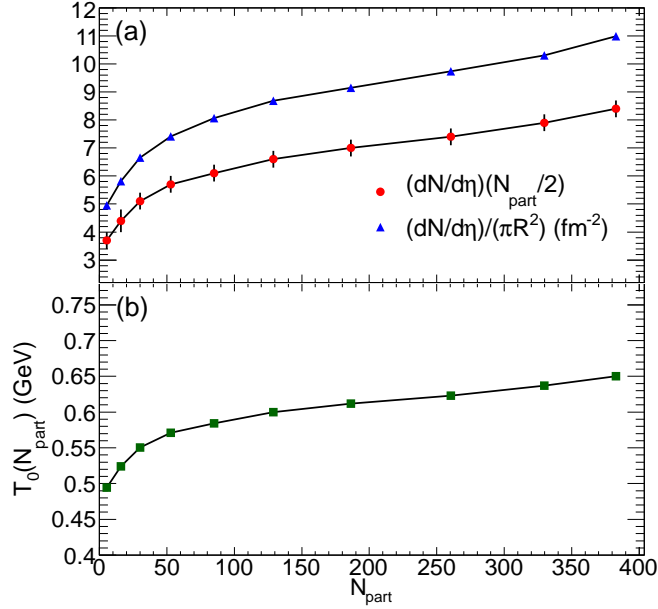}
\caption{
a) Measured $(dN/d\eta)/(N_{\rm part}/2)$ ~\cite{MULT} as a function 
of $N_{\rm part}$ along with the function $(dN/d\eta)/(\pi R^2)$. 
(b) The initial temperature obtained from measured multiplicity using Eq.~(\ref{Int2})
}
\label{fig:upsiRatio1}
\end{center}
\end{figure}
\begin{figure}
\begin{center}
  \includegraphics[width=0.48\textwidth]{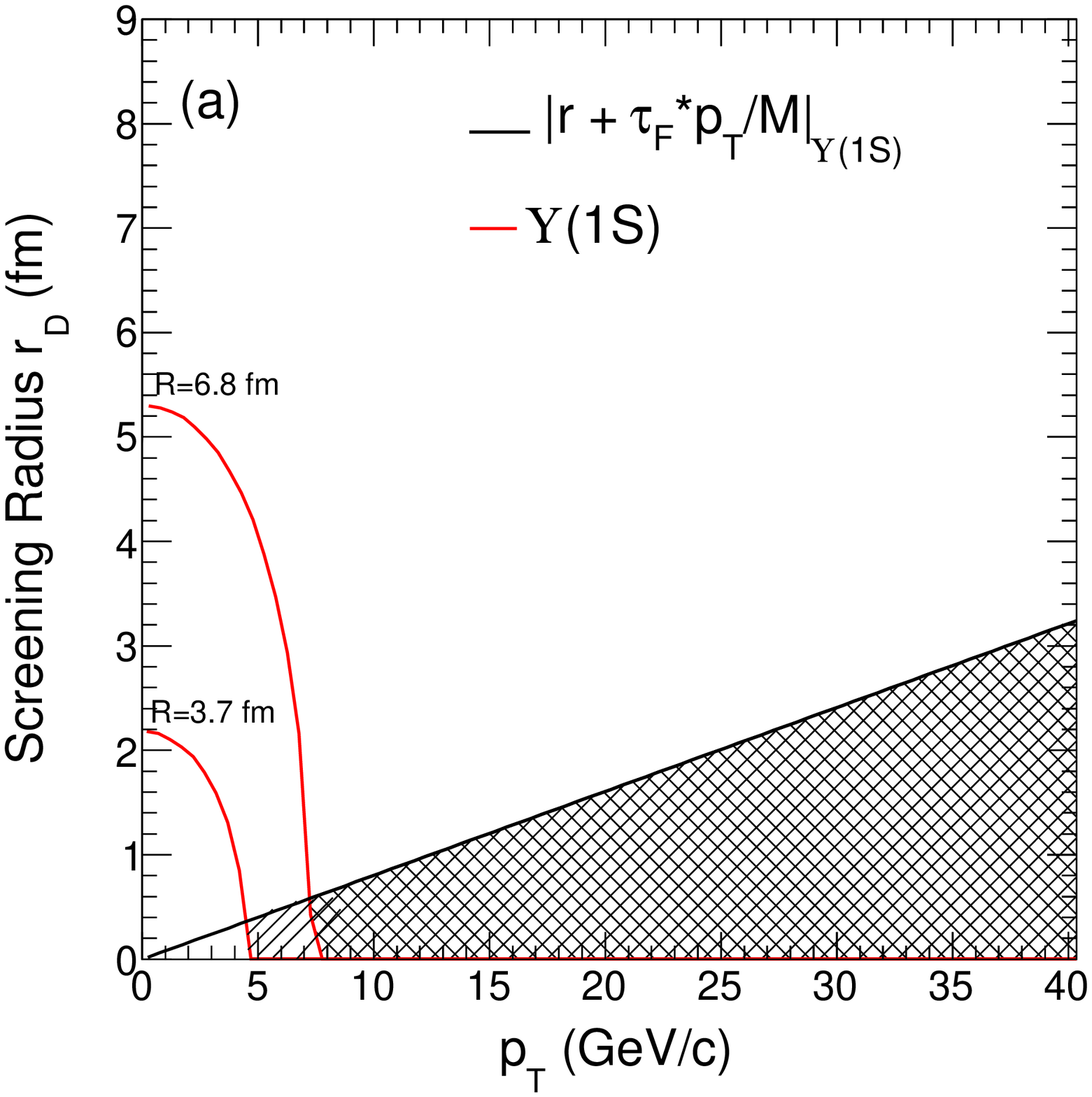}
  \includegraphics[width=0.48\textwidth]{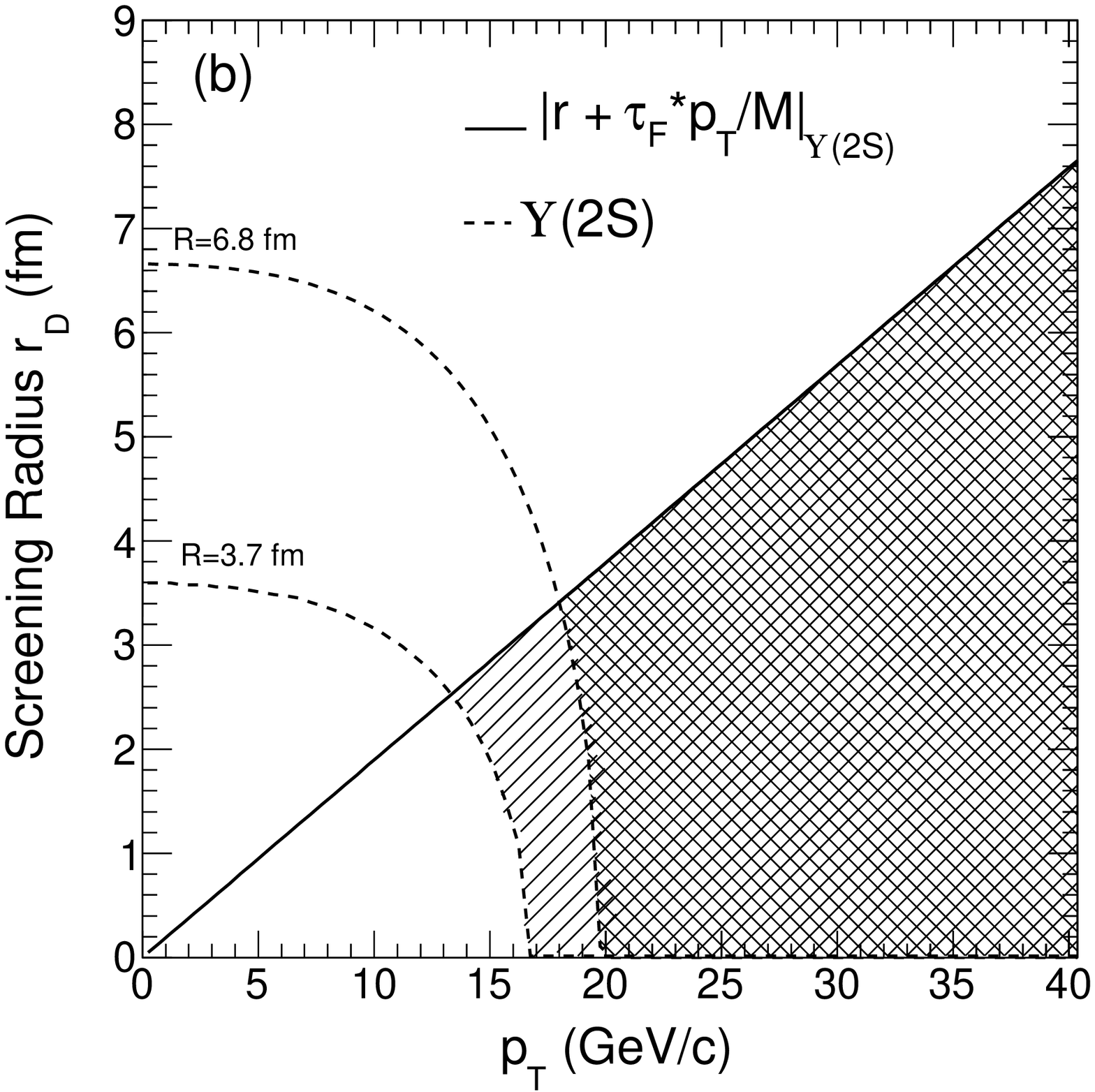} \\
\caption{ The screening radius $r_D$ (in fm) as a function of $p_T$ for $R=6.8$ fm 
(corresponding to head-on collisions) and $R=3.7$ fm (corresponding to minimum bias collisions)
for (a) $\Upsilon(1S)$ and (b) $\Upsilon(2S)$.
 The straight lines $|$ ${\rm \bf r} + {\tau_{F} {\rm \bf p_{T}} \over M}$ $|$
mark the distance a bottom quark pair (created at $r=0$) will travel before forming a bound state.
 The mesh region in both the figures marks the escape region for 
bottom quark pair in case of head-on collisions and total shaded (mesh+lines) region marks
the escape region in case of minimum bias collisions. }
\label{fig:upsiRatio2}
\end{center}
\end{figure}

  The nuclear modification factor, $R_{\rm AA}$ is obtained from survival probability taking into account 
the feed-down corrections as follows,

\begin{eqnarray}
 R_{\rm AA}(3S) &=& S(3S)  \nonumber \\
 R_{\rm AA}(2S) &=& f_1~S(2S) + f_2~S(3S)  \nonumber \\
 R_{\rm AA}(1S) &=& g_1 ~S(1S) + g_2~S(\chi_b(1P)) + g_3~S(2S) + g_4~S(3S)
\end{eqnarray}
The factors $f$'s and $g$'s are obtained from CDF measurement \cite{CDF}. The values 
of $g_{1}$, $g_{2}$, $g_{3}$ and $g_{4}$ are 0.509, 0.27, 0.107 and 0.113 respectively.
Here it is assumed that the survival probabilities of $\Upsilon(3S)$ and $\chi_{b}$(2P) 
are same and $g_4$ is their combined fraction. 
The values of $f_{1}$ and $f_{2}$ are taken as 0.50 guided by the work from Ref.~\cite{STR2}. 
%
\begin{figure*}
\begin{center}
\includegraphics[width=0.48\textwidth]{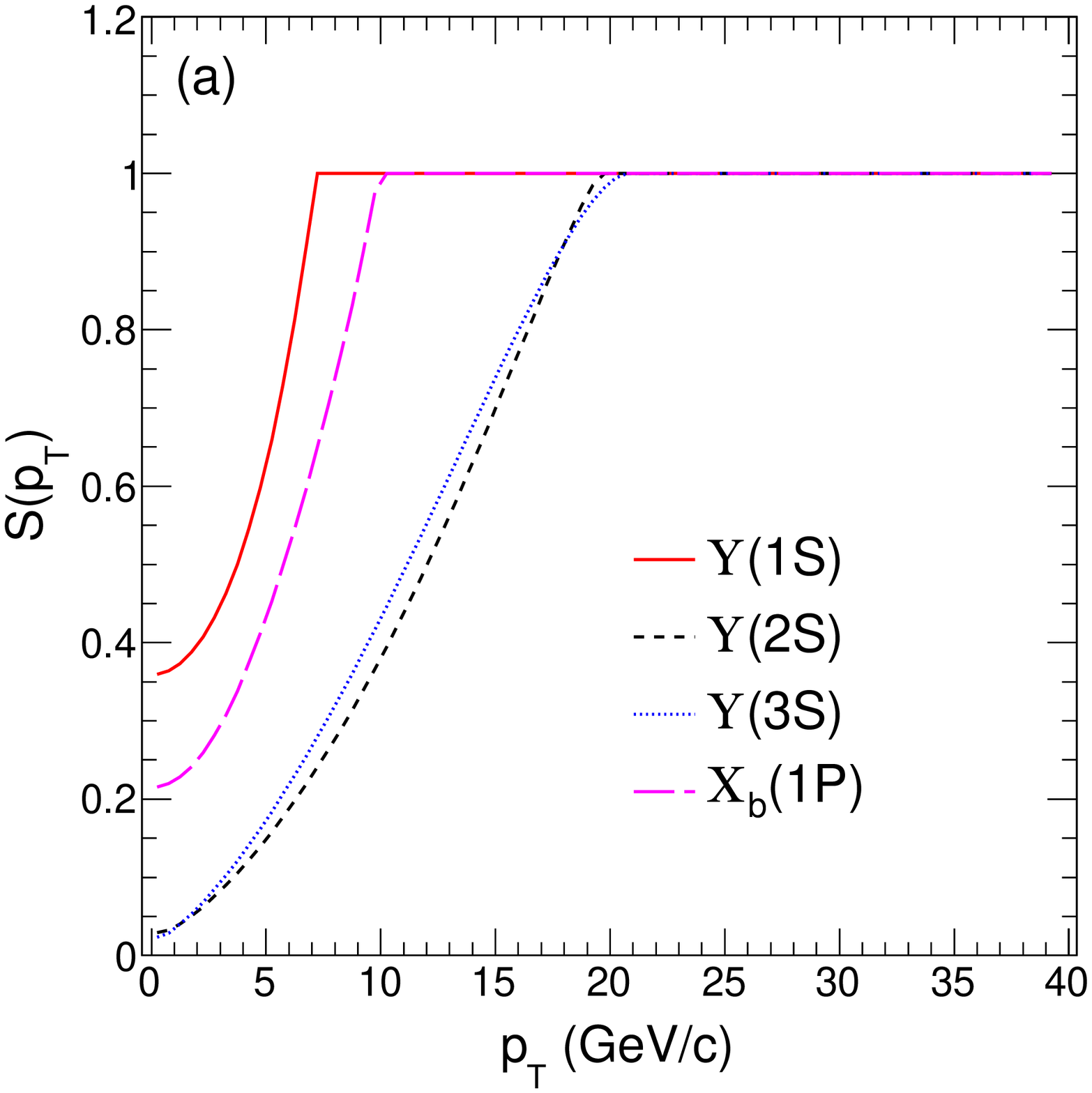} 
\includegraphics[width=0.48\textwidth]{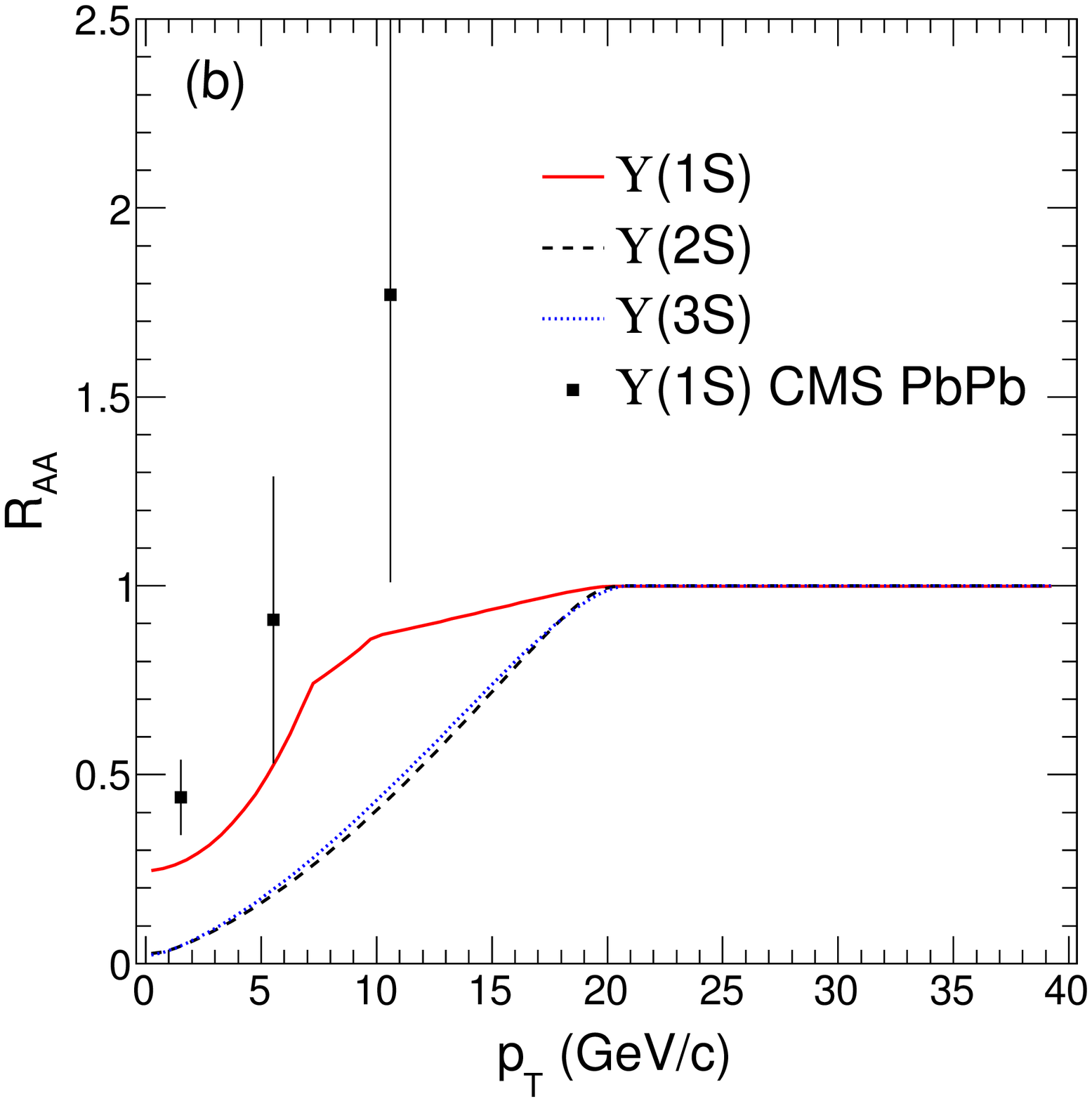}\\
\caption{ (a) The survival probability as a function of 
$p_{T}$ for $\Upsilon(1S)$, $\Upsilon(2S)$, $\Upsilon(3S)$ and $\chi_{b}(1P)$ for $R=3.7$ fm 
(corresponding to average $N_{\rm part}$ = 114 for minimum bias collisions). 
 (b) The nuclear modification factor 
for $\Upsilon(1S)$, $\Upsilon(2S)$ and $\Upsilon(3S)$ which is obtained from survival probabilities including 
feed down corrections.  The solid squares are $\Upsilon(1S)$ $R_{\rm AA}$ measured in the minimum 
bias PbPb collisions at $\sqrt{s_{NN}} = 2.76$ TeV by CMS experiment \cite{JCMS}. 
}
\label{fig:upsiRatio3}
\end{center}
\end{figure*}
\begin{figure*}
\begin{center}
\includegraphics[width=0.65\textwidth]{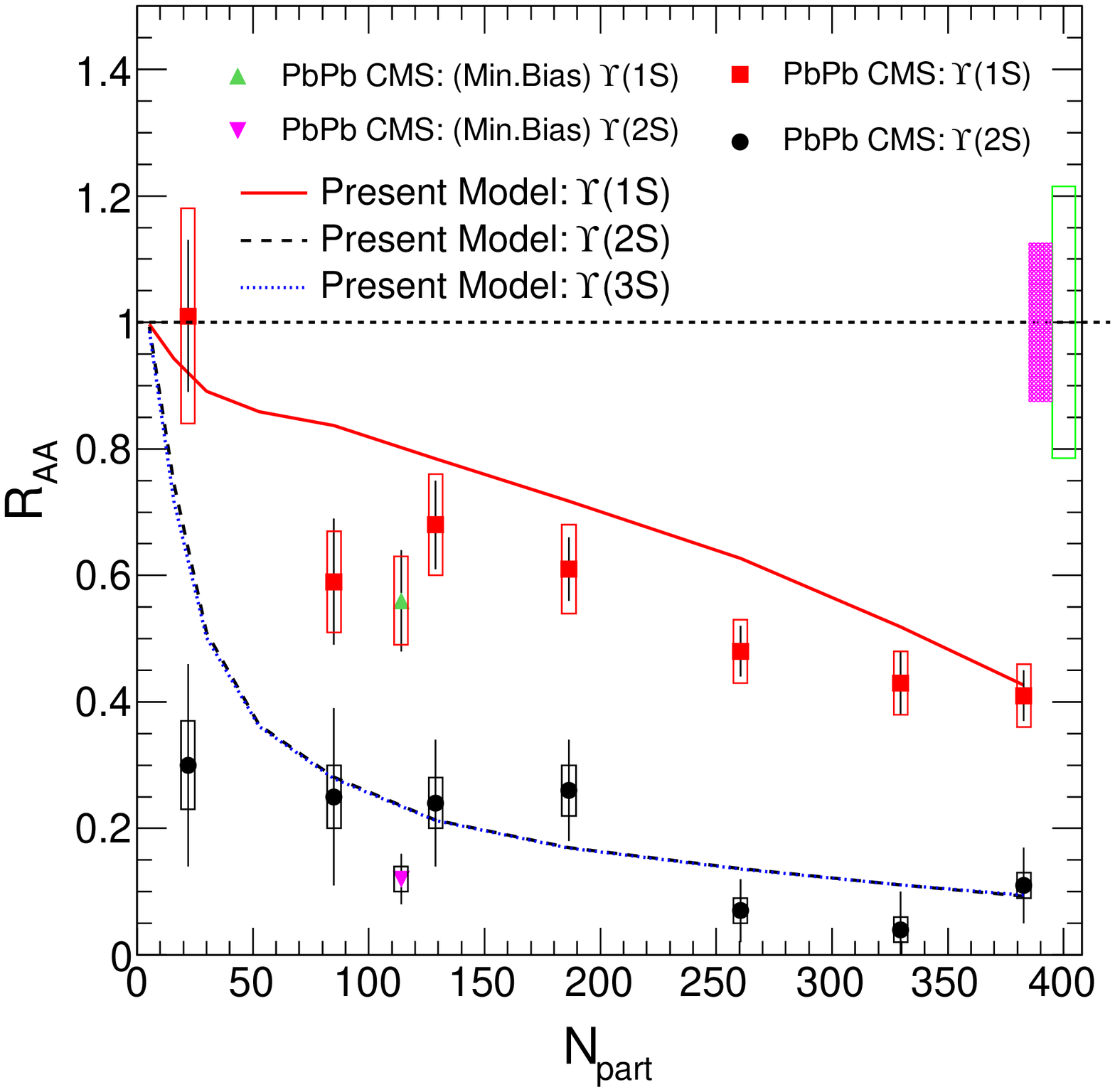} 
\caption{The nuclear modification factor, $R_{\rm AA}$
as a function of $N_{\rm part}$ for $\Upsilon(1S)$, $\Upsilon(2S)$ and $\Upsilon(3S)$. The solid squares and circles are 
measured $R_{\rm AA}$ by CMS experiment in PbPb collisions at $\sqrt{s_{\rm NN}}$ = 2.76 TeV \cite{CMSU2} for  
$\Upsilon(1S)$ and $\Upsilon(2S)$ respectively and solid triangles are the minimum bias data points. 
The boxes at unity are the common systematic uncertainties in pp luminosity 
measurement and the pp yield. The lines(solid for $\Upsilon(1S)$ , dashed for $\Upsilon(2S)$ and dotted for $\Upsilon(3S)$) 
represent the present model calculations. }
\label{fig:upsiRatio4}
\end{center}
\end{figure*}

\section{Results and discussions}
  Figure~\ref{fig:upsiRatio1} (a) shows measured $(dN/d\eta)/(N_{\rm part}/2)$ ~\cite{MULT} as a function 
of $N_{\rm part}$. The function $(dN/d\eta)/(\pi R^2)$ gives the multiplicity divided by transverse 
area obtained using Eq.(\ref{rnpart}). Figure~\ref{fig:upsiRatio1} (b) gives the initial temperature 
obtained from measured multiplicity using Eq.~(\ref{Int2}). Except in peripheral collisions, the initial 
temperature has weak dependence on centrality of collisions. 
Figure~\ref{fig:upsiRatio2} demonstrates working of the model. It shows
the screening radius $r_D$ (in fm) as a function of $p_T$ for $R=6.8$ fm 
(corresponding to head-on collisions) and $R=3.7$ fm (corresponding to minimum bias collisions)
for (a) $\Upsilon(1S)$ and (b) $\Upsilon(2S)$.
 The straight lines $|$ ${\rm \bf r} + {\tau_{F} {\rm \bf p_{T}} \over M}$ $|$
mark the distance a bottom quark pair (created at $r=0$) will travel before forming a bound state.
 The mesh region in both the figures marks the escape region for 
bottom quark pair in case of head-on collisions and total shaded (mesh+lines) region marks
the escape region in case of minimum bias collisions. 
 If $r$ is non-zero, the region where a bottomonium can escape screening, enlarges.

 Figure~\ref{fig:upsiRatio3} (a) shows the survival probability as a function of 
$p_{T}$ for $\Upsilon(1S)$, $\Upsilon(2S)$, $\Upsilon(3S)$ and $\chi_{b}(1P)$ for $R=3.7$ fm 
(corresponding to average $N_{\rm part}$ = 114 for minimum bias collisions). 
  The survival probability $S(p_{T})$ has a unique $p_T$ dependence decided by the 
$T_D$ and $\tau_{F}$ of each $\Upsilon$ state. 
  In general, the survival probabilities of resonance states increase with increasing $p_T$ 
and become unity at different $p_T$ for different states corresponding to complete survival. 
 Since $\Upsilon(1S)$ is expected to dissolve at a higher temperature it has more probability to survive
the plasma region even at lower p$_{T}$ as compared to the cases of other bottomonia states.
 The model gives very similar survival probabilities for $\Upsilon(2S)$ and $\Upsilon(3S)$.
This is due to the fact that $\Upsilon(3S)$ has large formation time even though its 
dissociation temperature is smaller in comparison to $\Upsilon(2S)$.
 Figure~\ref{fig:upsiRatio3} (b) shows the nuclear modification factor 
for $\Upsilon(1S)$, $\Upsilon(2S)$ and $\Upsilon(3S)$ which is obtained from survival probabilities 
including feed down corrections. 
 The solid squares are $\Upsilon(1S)$ $R_{\rm AA}$ measured in the minimum bias PbPb collisions at 
$\sqrt{s_{NN}} = 2.76$ TeV by CMS experiment \cite{JCMS}. 
 The model reproduces the trend of the $p_T$ dependence of low statistics 
measurements of $R_{\rm AA}$ from 2010 PbPb collisions by CMS.  

  Figure~\ref{fig:upsiRatio4} shows the nuclear modification factor, $R_{\rm AA}$
as a function of $N_{\rm part}$ for $\Upsilon(1S)$, $\Upsilon(2S)$ and $\Upsilon(3S)$. The solid squares and circles are 
measured $R_{\rm AA}$ by CMS experiment in PbPb collisions at $\sqrt{s_{\rm NN}}$ = 2.76 TeV \cite{CMSU2} for  
$\Upsilon(1S)$ and $\Upsilon(2S)$ respectively and solid triangles are the minimum bias data points. 
The lines(solid for $\Upsilon(1S)$, dashed for $\Upsilon(2S)$ and dotted for $\Upsilon(3S)$) represent 
the present model calculations. 
The common systematic uncertainties in pp luminosity measurement and the pp yield are 
represented by the boxes at unity. The model correctly reproduces the measured nuclear modification 
factors of both $\Upsilon(1S)$ and $\Upsilon(2S)$ for 
all centralities using the parameters given in the Table I. The 
survival probabilities for $\Upsilon(2S)$ and $\Upsilon(3S)$ are very similar.
\begin{figure}
\begin{center}
\includegraphics[width=0.65\textwidth]{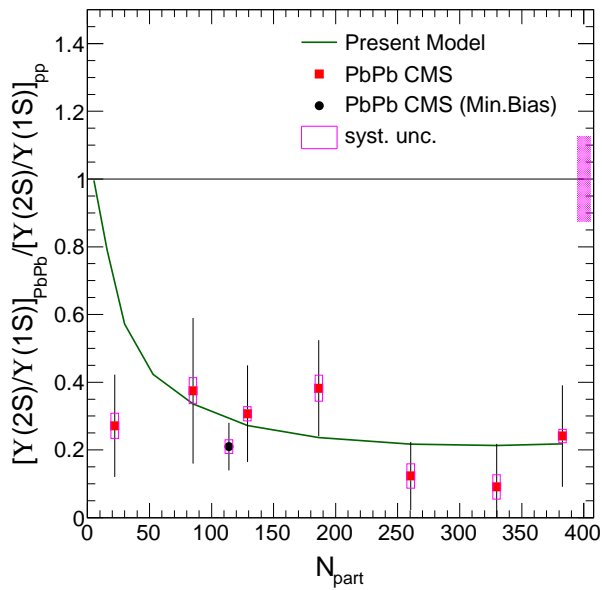}
\caption{Double ratio, $[\Upsilon(2S)/\Upsilon(1S)]_{PbPb}$/$[\Upsilon(2S)/\Upsilon(1S)]_{pp}$ as a 
function of $N_{\rm part}$ measured by CMS experiment \cite{CMSU2} 
along with the present calculation (solid line). The box at unity is
the common systematic uncertainty in the pp yield. 
}
\label{fig:upsiRatio5}
\end{center}
\end{figure}

 We also calculated the ratio of $R_{\rm AA}$ of $\Upsilon(2S)$ to that of $\Upsilon(1S)$ 
which is equivalent to the so called double ratio $[\Upsilon(2S)/\Upsilon(1S)]_{PbPb}$/$[\Upsilon(2S)/\Upsilon(1S)]_{pp}$.
The double ratio has the advantage that the effects such as initial-state nuclear effects and regeneration
which we ignore in our calculations are supposedly canceled out. 
 Figure~\ref{fig:upsiRatio5} shows the double ratio measured by CMS experiment \cite{CMSU2} 
along with the present calculation. The calculations reproduce the measured double ratio 
even for the most peripheral data point.

 The most important parameters in above study are formation time 
and dissociation temperatures of bottomonia states. There are reliable calculations of formation time 
obtained from zero temperature potential models which reproduce the bottomonia spectroscopy very well. 
Upper limits are available for dissociation temperatures which are obtained from potential models 
at finite temperature. We used slightly lower values of the dissociation temperature to get a good
description of the measured nuclear modification factors of $\Upsilon(1S)$ and $\Upsilon(2S)$. 
 The dynamics of the system is affected by the initial conditions which in the present calculations are 
fixed using measured charged particle multiplicity at LHC.
There can be suppression due to initial nuclear effects which we assume to be 
much smaller than that due to colour screening and hence are ignored in the present work. 
  The calculations of shadowing in PbPb show that it will effect the bottomoina yields by 
approximately 20 \% for most central collisions \cite{Shadow}. Thus, the dissociation temperatures 
obtained by us are still considered to be the upper limits. Conversly there are other views which say that  
$\Upsilon$ ground state is not much affected by the color screening 
up to the temperatures of $\sim 3-4T_C$ and regeneration of the states are not negligible at the LHC \cite{Rapp}.
  The bottom quark mass is 10 times higher than the temperature we 
are considering for the system and hence the regeneration effect can be safely ignored in calculating
nuclear modification for bottomonia. The uncertainties in the measurements of feed-down fractions 
would introduce uncertainties in the calculated nuclear modification factor. 
 Finally we mention that the uncertainties arising from the effects other than colour screening are small and supposedly 
will have little or no effect on the double ratio.

\section{Conclusions}
  In summary, we calculate the survival probabilities of $\Upsilon$ states and obtain the nuclear modification
factors due to colour screening in an expanding quark gluon plasma of finite lifetime and size produced 
during PbPb collisions $\sqrt{s_{NN}}=$ 2.76 TeV.
  The formation time and dissociation temperatures of bottomonia states 
obtained from potential models are used as input parameters in the model. 
We used slightly lower values of the dissociation temperatures to get a good
description of the measured nuclear modification factors of $\Upsilon(1S)$ and $\Upsilon(2S)$. 
  The model reproduces the centrality dependence of measured nuclear modification 
factors of $\Upsilon(1S)$ and $\Upsilon(2S)$ and the double ratio very well at $\sqrt{s_{\rm NN}}$ = 2.76 TeV.
  The trend of $p_T$ dependence of low statistics measurements of nuclear modification factor 
from 2010 PbPb collisions of CMS is reproduced as well.  
 The uncertainties arising from effects other than colour screening are assumed to be 
small and supposedly will have little or no effect on the double ratio calculations.
\section{Acknowledgement}
We thank Vineet Kumar, Ramona Vogt and CMS heavy ion colleagues for many fruitful discussion.
\end{document}